\documentclass[
    ,final            
  ]
  {aipproc}

\usepackage{amssymb}

\layoutstyle{6x9}

\begin{document}

\title{The coupling between pulsation and mass loss in massive stars}


\author{Rich Townsend}{
  address={Bartol Research Institute, Department of Physics \&
  Astronomy, University of Delaware, Newark, DE 19716, USA}
}

\begin{abstract}
To what extent can pulsational instabilities resolve the mass-loss
problem of massive stars?  How important is pulsation in structuring
and modulating the winds of these stars? What role does pulsation play
in redistributing angular momentum in massive stars? Although I cannot
offer answers to these questions, I hope at the very least to explain
how they come to be asked.
\end{abstract}

\maketitle


\section{Introduction}

What constitutes a `massive' star? For simplicity, I'm going to lump
together under this label all the stars that fall into the O and B
spectral types on the main sequence. I'm also going to include the
Wolf-Rayet (W-R) stars that are the late evolutionary states of
high-mass O stars; but not, however, the central stars of planetary
nebula, which can also show W-R spectra.

As will become apparent, massive stars are very different from their
cooler cousins that comprise the principal focus of this
conference. Throughout their lifetimes they shed a significant
fraction of their initial mass, and they often exhibit periodic or
episodic variability somehow associated with the large-amplitude
excitation of one or more pulsation modes. In the following section, I
briefly review the twin topics of mass-loss and pulsation; then, I
explore three different themes arising in the coupling between these
phenomena.

\section{Background}

\subsection{Mass loss from massive stars}

The evolutionary biologist \citet{Dob1964} once famously remarked that
``\emph{Nothing in biology makes sense except in the light of
evolution}''. With little massaging, we can convert this quote into an
equally-important truism for stellar astrophysicists: ``\emph{Nothing
in massive-star evolution makes sense except in the light of mass
loss}''. As I discuss below, to ignore mass loss when attempting to
model the evolutionary trajectory of a massive star is often so poor
an approximation that it isn't worth even considering.

First, however, I shall briefly review what we know
\emph{observationally} of the mass loss. In spectra, the appearance of
P Cygni-type line profiles, having symmetric emission superimposed
over blueshifted absorption \citep[see][and references
therein]{Rot1952}, is the unmistakable fingerprint of a circumstellar
envelope that is accelerating away from the star --- in other words, a
wind. Measurements of the violet edge of the absorption indicate wind
terminal velocities extending up to $v_{\infty} \sim 2,500\,{\rm km
s^{-1}}$ \citep[e.g.,][]{Pri1990}, significantly faster than found in
the Solar wind ($\sim 500\,{\rm km s^{-1}}$) or in the dusty winds of
AGB stars ($\sim 15\,{\rm km s^{-1}}$). Diagnostics based on H$\alpha$
and radio emission indicate corresponding mass-loss rates that reach
up to $\dot{M} \sim 10^{-5}\,{\rm M_{\odot} yr^{-1}}$
\citep[e.g.,][]{LamLei1993}; again, to place this value in context,
the Sun has $\dot{M} \sim 2 \times 10^{-14}\,{\rm M_{\odot} yr^{-1}}$
\citep{Woo2002}, around nine orders of magnitude smaller.

What causes these winds? In the case of the Sun, it is simply the gas
pressure in the hot ($\sim 10^{6}\,{\rm K}$) corona that drives the
outflow. However, to reach the terminal velocities seen in massive
stars would require temperatures reaching up to $T \sim (2,500\,{\rm
km s^{-1}})^{2} m_{p} /k \sim 5 \times 10^{8}\,{\rm K}$ (here, $m_{\rm
p}$ is the proton mass, and $k$ is Boltzmann's constant), which
clearly contradicts observations that reveal wind temperatures not too
different from that of the star. In fact, massive-star winds are
driven directly by radiation; UV continuum photons are scattered by
resonance lines of metallic ions, and in the process impart some of
their momentum to the ions. Through Coulomb coupling, these ions in
turn share the momentum with hydrogen and helium ions. If the net rate
of momentum deposition exceeds the local force of gravity, then a wind
outflow will ensue. This mechanism was first suggested by
\citet{LucSol1970}, but the full theory of radiatively driven winds
was developed in a groundbreaking paper by \citet*{Cas1975}. A lucid
introduction to this eponymous `CAK' theory can be found in the
extensive review by \citet{Owo2004a}, and this review also covers the
so-called `line-driven instability', that leads to self-seeded
structure in massive-star winds.

As I have already remarked, mass loss in a radiatively driven wind can
have a dramatic effect on the evolution of a star. As discussed by
\citet{ChiMae1986}, a high-mass ($\gtrsim 50\,{\rm M_{\odot}}$) star
in the absence of a wind will evolve to the red supergiant part of the
HRD; ignite helium; and remain there for the rest of its lifetime.
However, a very different picture emerges when the mass loss is
included. Figure 8 of \citet{Lan1994} shows a typical scenario: a
$60\,{\rm M_{\odot}}$ star makes only a brief evolutionary excursion
to the red side of the HRD, before returning to the blue, eventually
crossing over the ZAMS. This return arises because the star sheds its
hydrogen-rich envelope, revealing a hot, helium core showing
nucleosynthetic enrichments at first of nitrogen, and then of
carbon. The realisation that such wind-bared helium cores are none
other than Wolf-Rayet stars \citep[e.g.][and references
therein]{Con1983} provides the basis for unified narratives for
massive-star evolution \citep[e.g.,][]{Lan1994,SmiCon2007}.

Lest this all sound too straightforward, let me end this section by
highlighting a few significant unsolved problems (in accordance with
the overall theme of the conference!). Most significantly, there is
mounting evidence that the presence of wind clumping has led to
over-estimates in literature wind mass-loss rates \citep[see,
e.g.,][and references therein]{SmiOwo2006}. From diagnostics that are
insensitive to clumping, such as UV absorption lines \citep{Ful2006},
it seems likely that a reduction in $\dot{M}$ by a factor 3-10 is in
order. This raises an obvious problem: how then can we form a
(relatively low-mass) Wolf-Rayet star from a massive star, if the
latter only sheds a small fraction of its mass during its
hydrogen-burning phase? \citet{SmiOwo2006} address this issue by
suggesting that all stars above $\sim 40-50\,{\rm M_{\odot}}$ go
through a luminous blue variable (LBV) phase, during which they shed
copious amounts of mass in eruptions similar to the 19$^{\rm th}$
century outburst of $\eta$ Carinae. However, the mass-loss mechanism
involved in such eruptions remains unknown; I discuss one possibility
below.

Another area of significant uncertainty concerns rotation. Massive
stars are systematically rapid rotators; in a survey of 373 O and B
stars, \citet{How1997} found a distribution of projected equatorial
velocities $v \sin i$ with a peak at $\sim 100\,{\rm km s^{-1}}$, and
an extended tail reaching up to $\sim 400\,{\rm km s^{-1}}$. Recent
evolutionary calculations that incorporate rotation \citep[see][and
references therein]{MaeMey2000} reveal that in some cases massive
stars can pass through phases of super-critical rotation, during which
the centrifugal force exceeds the equatorial gravitational
force. These phases appear likely to be associated with significant
equatorial mass loss; but how to model this mass loss correctly
remains unclear \citep[e.g.,][]{Mey2006}. In this respect, the Be
stars --- characterized by the episodic formation of decretion disks,
perhaps due to critical rotation \citep{Tow2004} --- could be Rosetta
stones; but as I indicate below, pulsation also appears to be playing
a role in these enigmatic objects.

\subsection{Pulsation of massive stars}

\begin{figure}
  \includegraphics[height=.65\textheight]{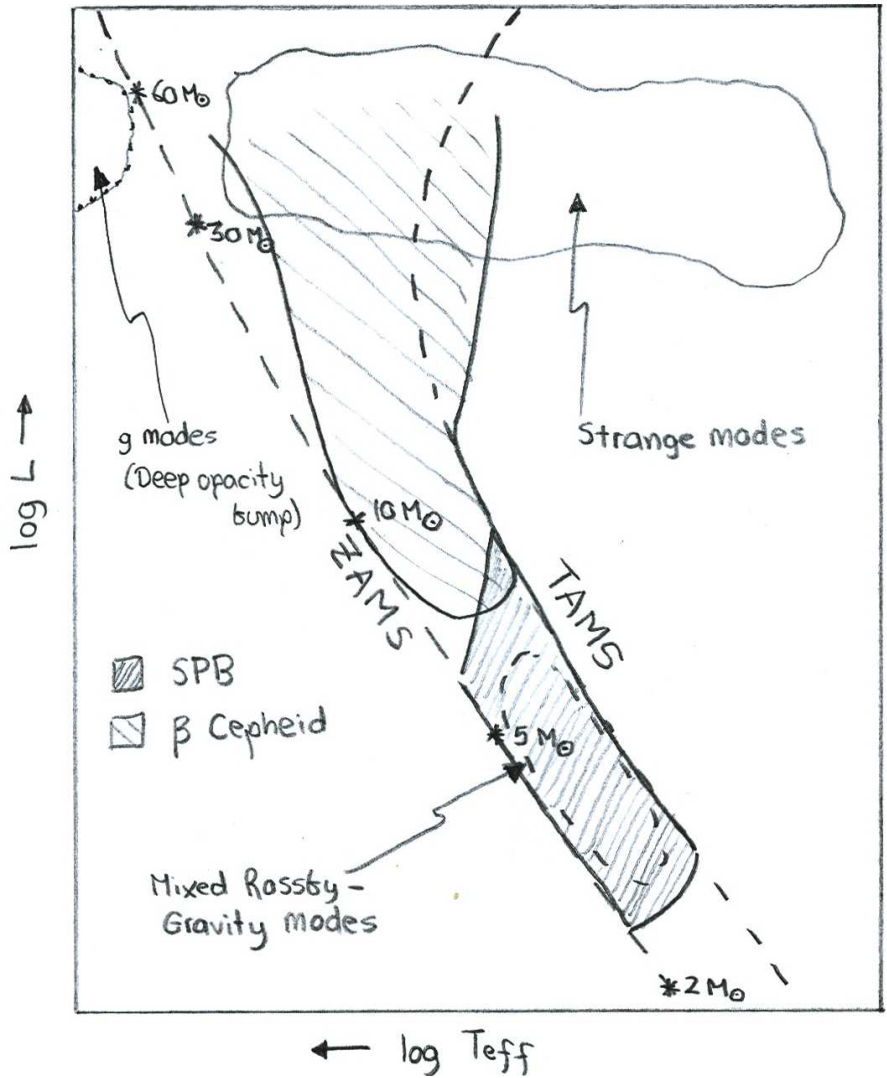}
  \caption{Instability strips in the upper part of the HRD
  \citep[after][his Fig.~3]{Pam1999}. The $\beta$ Cep and SPB instability strips
  associated with the iron-bump $\kappa$ mechanism are shown as the
  shaded regions; the approximate locations of unstable mixed
  Rossby-gravity modes, strange modes, and deep-opacity $g$ modes are
  also illustrated. The dotted lines indicate the zero-age (ZAMS) and
  terminal-age (TAMS) lines, and selected initial masses are indicated
  along the ZAMS.}
\label{fig:instability}
\end{figure}

To introduce the topic of massive-star pulsation, I present a potted
history of the field. Interest in this area of stellar astrophysics
grew initially from the strange case of $\beta$ Canis Majoris. Radial
velocity measurements of this star \citep{Str1950} indicate two
closely-spaced periods that are difficult to reconcile with either
binary motion or the radial pulsations seen in classical ($\delta$)
Cepheid stars. A seminal paper by \citet{Led1951} resolved this issue
by arguing that the star was undergoing nonradial oscillations in a
pair of rotationally split quadrupole modes. Although the formalism of
nonradial pulsation had been developed almost a century previously
\citep{Tho1863}, this was the first-ever example of a star oscillating
in this manner.

Subsequently, the Sun rather stole the limelight as the prime exemplar
of a nonradial oscillator \citep{Lei1962}. However, interest in
massive-star pulsation continued to grow with the discovery of other
stars like $\beta$ Canis Majoris, leading to the recognition of a
distinct class of variables: the `$\beta$ Cepheid' stars. (I confess
to being ignorant as to why $\beta$ Canis Majoris was demoted from its
status as the archetype.) \citet{Osa1971} brought a great deal of
quantitative rigour to the field, by reproducing the distinctive
line-profile variations (lpv) seen in time-series spectra of $\beta$
Cep stars. \citet{Smi1977} later discovered a distinct class of
early- to mid-B type pulsators, characterized by similar lpv but
exhibiting longer periods. During the 1980s this class was referred to
as `53 Per' stars, after the archetype; but \citet{Wae1991} recognised
that these stars are the spectroscopic counterparts of photometric
variables discovered by \citet{WaeRuf1985}. This led him to establish
a new `slowly pulsating B' (SPB) class, unifying the two groups.

In the 1990s, theorists finally caught up with these observational
strides, by uncovering the process(es) responsible for exciting
massive-star pulsations. Previously, \citet{Ste1978} had attempted to
invoke an opacity ('$\kappa$') mechanism based on helium second
ionisation, as successfully applied to classical Cepheids
\citep{Zhe1953}. Although ultimately unsuccessful, Stellingwerf was
prescient in surmising that opacity is the key. With the completion of
the OPAL and OP opacity calculations \citep{RogIgl1992,Sea1994}, it
became apparent that a new opacity peak at a temperature $\log T
\sim 5.3$ --- due to same M-shell transitions of iron and nickel
--- would destabilise $p$ modes (with periods on the order of hours)
in $\beta$ Cepheids, and $g$ modes (with periods on the order of days)
in SPB stars \citep[e.g.,][]{Cox1992,Dzi1993,DziPam1993}

Fig.~\ref{fig:instability} illustrates the instability strips in the
HRD diagram associated with this `iron-bump' $\kappa$ mechanism. As
demonstrated for instance by \citet{Pam1999}, the correspondence
between these strips and the observed positions of $\beta$ Cep and SPB
stars is extremely good. The figure also indicates the approximate
regions associated with other instabilities that can excite pulsations
in massive stars. For mid-B type stars, \citet{Sav2005} and
\citet{Tow2005} have independently shown that mixed Rossby-gravity
modes --- in which the restoring force on displaced fluid elements is
a combination of buoyancy and inertia --- are unstable due to the same
iron-bump $\kappa$ mechanism. Moreover, for Wolf-Rayet stars
\citet{TowMac2006} have demonstrated that $g$ modes are unstable due
to an opacity bump at $\log T \sim 6.25$, this time arising from
K-shell bound-free transitions of iron. Finally, a number of authors
have investigated so-called `strange' instabilities, that cause
violent radial and nonradial pulsations in objects characterized by
large luminosity-to-mass ratios \citep[see][and references
therein]{Sai1998}. These strange instabilities are discussed further
in the following section.

\section{Themes}

\subsection{Pulsation-driven mass loss}

What limits the mass of the star? There is strong observational
evidence that the initial mass function (IMF) in the Milky Way is
truncated at the high-mass end, with no single star exceeding
$150\,{\rm M_{\odot}}$ \citep{Fig2005}. However, the underlying cause
of this upper mass limit remains a puzzle. Historically, it was
thought that radial pulsations driven by the $\epsilon$ mechanism ---
an instability due to the large temperature exponent of CNO burning
--- would disrupt any star having a mass above $\sim 60\,{\rm
M_{\odot}}$ \citep{SchHar1959}. This expectation was lent support by
\citet{App1970}, although his calculations suggested a higher
theoretical limit for complete disruption, $\gtrsim 300\,{\rm
M_{\odot}}$. However, subsequent investigations by
\citet{Pap1973a,Pap1973b} gave the contrary result that \emph{no} mass
loss is expected to be driven by $\epsilon$-mechanism
pulsation. Interest in the issue then subsided, overshadowed perhaps
by the then-rapid advances in understanding \emph{wind} mass loss.

The situation changed two decades later, when \citet{GlaKir1993}
discovered strange-mode instabilities whose theoretical location in
the HRD appeared to be correlated with the observational
Humphreys-Davidson (HD) limit \citep{HumDav1979}. To explain why this
is significant, I shall first attempt to shed some light on the
confusing topic of strange modes and strange instabilities. Amongst
the various differing definitions suggested for strange modes, my
preference is for the one given by \citet{Sai1998}: ``\emph{As strange
modes we identify those eigenfrequency branches that behave
differently from those that change only very slowly under the change
of a control parameter}''. Typically, the `control parameter' that
these authors refer to is the age of a star; and thus a strange mode
can be thought of as one whose frequency changes rapidly as the star
evolves, in contrast to the slower variation in the frequencies of
ordinary modes. 

A good example of such behaviour can be seen in the top-left panel of
Fig. 2 of \citet{Sai1998}. It originates because of the way in which
the modes are trapped in the star. For ordinary $p$ modes, the
trapping depends on the variation in adiabatic sound speed $c_{\rm
ad}$ throughout the star; steep gradients in $c_{\rm ad}$ tend to
reflect acoustic waves, establishing the boundaries of trapping
zones. \citet{GlaKir1993} argued that strange modes are acoustic waves
that trapped in the outer layers of the star by steep gradients of
some appropriately defined sound speed $c$. In some cases $c = c_{\rm
ad}$, meaning that the strange modes are essentially ordinary $p$
modes; these are what \citet{Sai1998} term `adiabatic strange
modes'. In other cases, non-adiabatic effects can mean that $c$
departs significantly from $c_{\rm ad}$, and there is no obvious
relation between strange and $p$ modes, apart from the fact that both
originate in acoustic waves. In all cases, however, the quality that
makes a mode \emph{strange} is that it is trapped in the outer layers
of the star; these layers tend to change rapidly as the star evolves,
giving rise to correspondingly rapid changes in the mode's frequency.

This brings me back to the relationship between strange modes and mass
loss. Being confined to the surface layers of a star means that the
growth timescale $\tau$ of these modes --- when some suitable driving
mechanism is operative --- can be on the same order as the dynamical
timescale $\tau_{\rm dyn}$ of the star. In practice, the driving
mechanism could be a He-ionisation $\kappa$ mechanism
\citep[e.g.,][]{GlaKir1993}; it could be the iron-bump $\kappa$
mechanism \citep[e.g.,][]{Kir1993}; or it could be the strange
instability, which I discuss further below. In each case, as $\tau
\rightarrow \tau_{\rm dyn}$ the prospect of significant hydrodynamical
mass loss arises. Determining the ultimate outcome involves following
the instability into the nonlinear regime; this is an extremely
challenging radiation-hydrodynamics problem, but
\citet{Gro2003,Gro2004,Gro2005} appear to be making good initial
progress.

Let me turn now to the strange instability --- which although related
to strange modes, should not be confused with them. The strange
instability arises in circumstances where the gas pressure $p_{\rm
gas}$ is small compared to the radiation pressure $p_{\rm
rad}$. Ordinarily, the converse is true, and the total pressure $p$
can be approximated by the ideal gas pressure,
\[
p \approx p_{\rm gas} =  \frac{\rho k T}{\mu}.
\]
This means that perturbations to the density $\rho$ are proportional to
perturbations to the pressure,
\[
\delta \rho \propto \delta p,
\]
and disturbances propagate as acoustic waves. When $p_{\rm gas} \ll
p_{\rm rad}$, however, the relation between density and pressure
perturbations should be obtained from the radiative diffusion
equation,
\[
F_{\rm rad} = \frac{1}{3 \kappa \rho} \nabla p_{\rm rad} \approx
\frac{1}{3 \kappa \rho} \nabla p.
\]
Assuming that the radiative flux $F_{\rm rad}$ remains constant (which
is appropriate in the envelopes of very massive stars, because the
radiative relaxation time is so short), then density perturbations are
governed by
\[
\delta \rho \propto \nabla \delta p.
\]
The gradient operator on the right-hand side introduces a
quarter-cycle phase shift in the dispersion relation governing
disturbances, and instead of obtaining wave solutions we find
exponential growth corresponding to instability. 

The foregoing discussion gives the simplest-possible view of the
strange instability, and glosses over many important issues; for a
rigorous treatment, refer to the discussion \S4.2 of
\citet{Sai1998}. Nevertheless, this basic analysis captures the
fundamental character of the strange instability --- namely, that it
is driven directly by radiation pressure, rather than relying on a
Carnot-cycle heat engine as found in `ordinary' instabilities such as
the $\kappa$ and $\epsilon$ mechanisms.

\subsection{Pulsation and stellar winds}

At amplitudes that are too small to eject mass directly, pulsations
can still play an important role in modulating mass-loss from a
massive star, by coupling to the star's radiatively driven
wind. Observationally, their is persuasive evidence that such coupling
takes place. Extended time-series spectra obtained using \emph{IUE}
reveal that the UV P Cygni absorption lines of many O and B stars
exhibit discrete absorption components (DACs), that migrate from red
to blue in a cyclical fashion \citep[see,
e.g.,][]{Pri1988,Pri1992,How1993}. At the same time, a survey of
optical line profiles in these stars by \citet{Ful1996} found
statistically significant lpv in 77\% (23/30) of their sample. It is
natural to speculate that these two types of variability must be
causally linked; but the direct evidence confirming such a
`photospheric connection' has proven quite difficult to come by.

The impasse appears to have been at least weakened by \citet{Kau2006},
who have conducted a detailed analysis of the optical lpv of the B0
supergiant HD~64760. This star shows some of the most dramatic UV
variations of any massive star, consisting of the episodic appearance
of migrating DACs superimposed over periodic (1.2\,d and 2.4\,d)
modulations in the absorption troughs of resonance line profiles
\citep{Mas1995}. Building on an idea first proposed by
\citet{Mul1986}, \citet{Ful1997} proposed that the periodic
modulations are due to the passage of corotating interaction regions
(CIRs) across the face of the star. As shown in earlier hydrodynamical
simulations by \citet{CraOwo1996}, these CIRs can be formed by
collisions between fast and slow wind streams that are rooted in flow
inhomogeneities at the stellar surface.

\citet{Kau2006} demonstrated a pulsational origin for these surface
structures, by showing that beating between three closely-spaced,
high-order ($\ell=6-10$) modes would lead to a 6.8\,d period that is
directly observed in wind-sensitive lines such as H$\alpha$. This
\emph{almost} amounts to a confirmation of a photospheric connection;
but a stumbling block remains the mismatch between the 6.8\,d period
in the photosphere and wind base, and the 1.2\,d/2.4\,d modulation
period seen in the UV absorption lines formed further out into the
wind. Indeed, the 6.8\,d period seems to correspond better with the
typical recurrence time of the DACs superimposed over the periodic UV
modulations.

Turning now to theoretical issues, progress has been slow in
understanding how pulsation and wind mass-loss can interact with one
another. The present author \citep{Tow2000a,Tow2000b} examined the
possibility that pulsation waves are not completely reflected at the
stellar surface, and instead leak through into the wind, possibly
seeding structure at the wind base that evolves into a CIR. In
principle this mechanism could work, but in practise the frequency of
the pulsation modes typically seen in massive stars fall in between
the twin critical frequencies $\omega_{\rm c_{1}},\omega_{\rm c_{2}}$
of the photosphere, meaning that in all cases complete wave reflection
occurs.

Significantly, however, this analysis did not account for the prior
existence of a wind. The expected impact of a wind is twofold. First,
it leads to a shallower stratification in the surface layers of a
star, with the density falloff transitioning from ${\rm e}^{-r}$ to
$r^{-2}$; this tends to make it more difficult for wave reflection to
occur. Second, the mean flow associated with a wind modifies the wave
propagation, meaning that even for frequencies between the formal
critical frequencies, complete reflection cannot occur. To my
knowledge, this latter effect has only been studied locally, and in
the simple case of an isothermal atmosphere \citep[see][]{Cra1996}; no
attempt has been made to include it into a global pulsation
code. Certainly, there is much scope for progress here.

To complete the discussion, I shall say a few words on small-scale
atmospheric structure due to pulsation. In the majority of pulsating
massive stars, we only detect a handful of modes --- in spite of the
fact that a linear analysis suggests hundreds if not thousands should
be unstable. Are these modes in fact damped? Or is it rather that
their amplitudes are too small for present-day instrumentation to pick
up? The \emph{COROT} mission and similar endeavours will help resolve
this issue, by pushing down detection thresholds to the micromagnitude
level. In the meantime, we can ask ourselves whether we already see
the signatures of many small-amplitude pulsation modes in massive
stars, under the twin guises of microturbulence and macroturbulence?
Both of these `phenomena' arise from the inability to fit photospheric
line profiles without assuming an additional source of Doppler
broadening; they only differ in the scale of the velocity structures
they assume, with micro (macro) being smaller (larger) than photon
mean free paths. To obtain consistent fits to Helium lines of O-type
supergiants, \citet{SmiHow1998} had to assume microturbulent
velocities on the order of $\sim 15\,{\rm km s^{-1}}$. Likewise,
\citet{Rya2002} found that macroturbulent velocities on the order of
$\sim 50\,{\rm km s^{-1}}$ were necessary to obtain acceptable fits to
the line profiles of B-type supergiants.

The origins of micro- and macroturbulence are of interest not only to
spectroscopists. A recent paper by \citet{Luc2007} has highlighted the
idea that the mass-loss rate in a radiatively driven wind can be sensitive
to the degree of microturbulence in the subsonic parts of the
outflow. Moreover, macroturbulence may impact observational $\dot{M}$
measurements, by modulating the wind clumping discussed previously. In
the absence of photospheric perturbations, the line-driven instability
\citep{FelOwo1998} causes small-scale wind clumping to arise
spontaneously at a few tenths of a stellar radius above the stellar
surface. However, the character and extent of the clumping may change
dramatically in the presence of macroturbulent/pulsation velocity
fields in the photosphere.

Related to this discussion is the issue of super-Eddington mass
loss. In a star whose surface layers are formally above the Eddington
limit (so that the electron-scattering radiative force exceeds
gravity), the resulting continuum-driven wind will be characterized by
a mass-loss rate far in excess of the $\dot{M}$ typical to line-driven
winds. \citet{SmiOwo2006} argue that the LBV eruptions they invoke, to
allow the transition to a Wolf-Rayet star, take the form of
continuum-driven winds. As discussed by \citet{Owo2004b}, steady
continuum driving requires some way to modulate the
electron-scattering opacity, and one way to do this is to invoke an
instability-driven porous atmosphere \citep{Sha2001}. It is not yet
clear how Shaviv's instability fits into a pulsational framework, but
it seems likely to be related to the strange instability discussed
previously.

\subsection{Pulsation and rotation}

The Be stars \citep[see][]{PorRiv2003} appear to be an ideal
laboratory for learning about the interplay between pulsation,
rotation, and mass loss. For a long time, the study of these objects
was dominated by arguments over whether their lpv and photometric
variations were due to pulsations or to rotational modulation of spots
and circumstellar structures \citep[e.g.,][]{BaaBal1994}. However, the
detection of multiple periods in the lpv of the B2e star $\mu$ Cen by
\citep{Riv1998}, and the successful modeling of these lpv by
\citet{Riv2001}, gave considerable weight to the pulsational
interpretation. A subsequent investigation by \citet{Riv2003}, using
the same \textsc{bruce/kylie} modeling codes as before
\citep[see][]{Tow1997}, revealed that the photospheric lpv of the
majority of variable Be stars can be attributed to nonradial pulsation
in retrograde, $\ell=m=2$ modes.

An interesting result from the analysis of $\mu$ Cen is that the
outbursts of the star, during which additional H$\alpha$ emission
appears, seem to be correlated with beating of the pulsation
modes. This appears to support an idea advanced by \citet{And1986} and
\citet{Osa1986}, that the Be phenomenon arises when pulsation waves
deposit sufficient angular momentum in the stellar surface layers for
these layers to reach critical rotation. The resulting lifting of
material into orbit then forms a viscous decretion disk, as envisaged
by \citet{Lee1991}. Hydrodynamical simulations by \citet{Owo2005} seem
to support such a process, but only for modes that are propagating in
the prograde direction. Unfortunately, as mentioned above the
preponderance of Be stars --- including $\mu$ Cen --- instead exhibit
retrograde pulsation. One way out of this seeming contradiction has
been suggested by \citet{Tow2005}: the mixed Rossby-gravity modes that
are unstable in B-type stars show a retrograde phase velocity, but a
prograde group velocity. Thus, they be able simultaneously to satisfy
the observational and theoretical constraints.

Perhaps, however, the problem is that the hydrodynamical simulations
by \citet{Owo2005} are not able to capture the full physics of the
situation. The net deposition of angular momentum in the surface
layers requires \emph{either} that prograde pulsation modes are
dissipated there, \emph{or} (in a recoil effect) that retrograde modes
be excited there. The latter scenario is what arises in SPB and
$\beta$ Cep pulsators; work functions for unstable modes in these
stars \citep[e.g.,][their Fig.~1]{Dzi1993} reveal an outer excitation
region associated with the iron-bump $\kappa$ mechanism, and an inner
dissipation region associated with radiative damping. So, in fact it
seems more likely that retrograde modes are required to spin up the
surface layers of these stars. That the simulations by \citet{Owo2005}
did not confirm this result, may be due to the boundary conditions
that were adopted.

Continuing with this theme of angular momentum transport, the
proximity of the excitation and dissipation regions in $\beta$ Cep and
SPB stars suggests that we should naturally expect a shear layer to
develop between them. This layer represents a source of free energy
that could, for instance, be tapped into to generate a magnetic
field. This is a particularly intriguing possibility, especially in
light of the reported detection of magnetic fields in 11 out of a
sample of 25 SPB stars \citep{Hub2007}. Further calculations are
needed to gauge the magnitude of the shear, which will be set by the
strength of the diffusive processes that act in competition with the
wave transport. At this early stage, I shall only remark that the
typical angular momentum luminosities due to massive-star pulsations
are expected to be orders-of-magnitude greater than found in
solar-type stars (see, e.g., Charbonnel, these proceedings); this is
simply because of the much-higher amplitudes of the instability-driven
modes in the former, as compared to the stochastically excited modes
in the latter. Are we therefore missing an important ingredient in our
understanding of the rotational evolution of massive stars?

\section{Summary}

To summarize my discussion, I restate the questions that I posed in
the abstract:

\begin{itemize}
\item To what extent can pulsational instabilities 
  resolve the mass-loss problem of massive stars?
\item How important is pulsation in structuring and modulating the
  winds of these stars?
\item What role does pulsation play in redistributing angular momentum
  in these stars?
\end{itemize}

The first question is relatively new, arising both from the discovery
of strange modes and strange instabilities, and from the recent
realisation that wind mass-loss rates are too small for Wolf-Rayet
stars to form. However, the latter two questions extend back at least
two decades, to a 1985 workshop held at the Joint Institute for
Laboratory Astrophysics (Boulder, Colorado). In a fascinating series
of connected papers, under the main title ``\emph{The Connection
Between Nonradial Pulsations and Stellar Winds in Massive Stars}'',
\citet{Abb1986} and other authors reviewed many of the topics I have
discussed in this contribution. That these questions still remain as
Unsolved Problems in Stellar Astrophysics is frustrating, for --- as I
hope I have been able to convey --- the need to understand the
coupling between pulsation and mass loss is even more pressing today
than it was all those years ago.


\begin{theacknowledgments}
I acknowledge support from NASA \emph{Long Term Space Astrophysics}
grant NNG05GC36G.
\end{theacknowledgments}



\bibliographystyle{mn2e}   

\bibliography{townsend}

\IfFileExists{\jobname.bbl}{}
 {\typeout{}
  \typeout{******************************************}
  \typeout{** Please run "bibtex \jobname" to obtain}
  \typeout{** the bibliography and then re-run LaTeX}
  \typeout{** twice to fix the references!}
  \typeout{******************************************}
  \typeout{}
 }

\end{document}